\documentclass[%
preprint,
amsmath,
amssymb,
aps,
longbibliography,
]{revtex4-2}
\usepackage[colorlinks=true, citecolor=blue, urlcolor=blue, linkcolor=blue]{hyperref}
\usepackage{graphicx}% Include figure files
\usepackage{dcolumn}% Align table columns on decimal point
\usepackage{bm}% bold math
\usepackage{xcolor}
\usepackage{graphicx}

\begin{document}

%\preprint{APS/123-QED}

\title{Optical Dipole Trap for Hg Atoms}% Force line breaks with \\

\author{I.~Nandi}
\author{A.~Sahu}
\author{M.~Veis}
\author{A.~Linek}
\thanks{Present address: EmpiriQa sp. z o.o., Jurija Gagarina 7/47, PL-87-100 Toru\'n, Poland}
%\altaffiliation{asd}
\author{R.~Ciury\l{}o}
\author{M.~Witkowski}
\email{marcin\_w@umk.pl}
\affiliation{Institute of Physics, Faculty of Physics, Astronomy and Informatics, Nicolaus Copernicus University, Grudzi\c{a}dzka 5, PL-87-100 Toru\'n, Poland.}%Lines break automatically or can be forced with \\

\date{\today}% It is always \today, today,

\begin{abstract}
We report optical dipole trapping of a system with low polarizability by demonstrating the first realization of a single-beam optical dipole trap (ODT) for ultracold mercury atoms, overcoming the challenge posed by its weak optical trapping potential. We demonstrate the transfer of six naturally abundant Hg isotopes from a magneto-optical trap into the ODT. Dense ultracold Hg samples enable studies of isotope-dependent collisions, photoassociation, searches for new physics beyond the Standard Model, and open a path toward quantum degeneracy.
%We report the first realization of an optical dipole trap (ODT) for ultracold mercury atoms, overcoming the challenge posed by the exceptionally low polarizability of Hg and the resulting weak optical trapping potential. We demonstrate the transfer of six naturally abundant Hg isotopes from a magneto-optical trap (MOT) into the ODT and characterize its loading dynamics, trap depth, and lifetime. Confinement of dense samples of ultracold Hg opens new opportunities for studies of isotope-dependent collisions and photoassociation, and provides a route toward quantum degeneracy.
%We demonstrate the transfer of atoms from a magneto-optical trap (MOT) into the ODT and characterize its loading dynamics, trap depth, and lifetime. 
%The ability to confine dense samples of ultracold mercury opens new opportunities for studies of isotope-dependent collisions, photoassociation, and provides a route toward quantum degeneracy of Hg.
 %other interaction-driven phenomena requiring high atomic densities.

\end{abstract}
\maketitle
%\maketitle

%\section{Introduction}

%Atomic displacement by resonant light was observed \cite{Ashkin1970a} through the action of a pronounced scattering force with the possibility of trapping atoms \cite{Ashkin1970b}. 
%Various cooling and trapping techniques have been developed for studying cold atomic gases, with the optical dipole trap emerging as one of the most widely adopted and well-established methods~\cite{Metcalf1999,Chu1986,Grimm2000}.
%The ODT provides a highly localized and conservative potential for trapping atoms, typically allowing for atomic densities up to three orders of magnitude higher compared to a magneto-optical trap, all without the need for magnetic fields.

%The ODT has proven invaluable for experiments such as photo-association spectroscopy~\cite{Fatemi2000, Jones2006, Cheng2010} 
%absorption spectroscopy with in-situ imaging??, 
%and Electromagnetically Induced Transparency~\cite{Kampschulte2014}. By producing dense atomic samples, the ODT facilitates the achievement of ultra-cold conditions, which are essential for exploring quantum phenomena like Bose-Einstein condensation~\cite{Barret2001}.
\textit{Introduction}---Optical dipole traps (ODTs) have become a standard tool for producing dense ultracold atomic samples and have enabled a broad range of experiments in atomic physics, including photoassociation spectroscopy and ultracold collision studies~\cite{Fatemi2000,Jones2006,Tojo2006}, quantum-degenerate gases~\cite{Barret2001,Stellmer2013}, and precision studies of quantum many-body systems~\cite{Gustavson2001,Zwierlein2006,Bloch2008,Chen2024}.
%Optical dipole traps (ODTs) have become a standard tool for producing dense ultracold atomic samples and form the basis of numerous experiments in atomic physics, including photoassociation spectroscopy, electromagnetically induced transparency, and quantum-degenerate gases~\cite{Fatemi2000, Jones2006, Cheng2010, Kampschulte2014, Barret2001}~\cite{Fukuhara2009}. 
Unlike magneto-optical traps (MOTs), ODTs provide conservative confinement free of magnetic fields, enabling atomic densities several orders of magnitude higher.
The trapping potential in an ODT arises from the interaction of off-resonant laser light with the induced atomic dipole moment. For red-detuned light, atoms are attracted to regions of maximum intensity, typically created by a tightly focused laser beam. Since ODTs provide relatively shallow conservative confinement that is largely independent of the magnetic sublevel, atoms must first be precooled, most commonly in a MOT.

The dipole force on an atom is proportional to its dynamic polarizability
%The strength of the dipole force $F_{dip}$ acting on atoms depends on their polarizability 
$\alpha(\omega_{\rm{dip}})$, and is given by~\cite{Grimm2000}
\begin{equation}
F_{\rm{dip}}\left(\bf r\right)=-\nabla U_{\rm{dip}}\left(\bf r\right)=\frac{1}{2\epsilon_0c}\text{Re}\left(\alpha\left(\omega_{\rm{dip}}\right)\right)\nabla I\left(\bf r\right),
\end{equation}
\noindent where $U_{\rm{dip}}\left({\bf r}\right)$ is the optical dipole potential experienced by the atoms, $I(\bf r)$ is the intensity of the dipole beam with angular frequency $\omega_{\mathrm{dip}}=2\pi c/\lambda_{\mathrm{dip}}$, $\lambda_{\mathrm{dip}}$ is the dipole laser wavelength,  $\epsilon_0$ is the vacuum permittivity, and $c$ is the speed of light in vacuum. Consequently, optical trapping becomes particularly challenging for atoms and molecules with low polarizability, such as Hg~\cite{Romalis1999} or H$_2$~\cite{Ubachs2025}. Among low-polarizability species, mercury is particularly attractive due to its distinctive properties, including a high atomic number, a rich isotopic structure, and narrow optical transitions, making it ideal for a variety of precision measurements and fundamental physics experiments~\cite{Graner2016,Safronova2018}. 
\begin{table}[ht]
    \centering
    \begin{tabular}{|c|c|c|c|c|}
        \hline
        \rule{0pt}{3ex}Z & Atom & State & $\alpha_s$  \\ \hline
        \rule{0pt}{3ex}3 & Li & $\mathrm{2s^1}\,\mathrm{^2S_{1/2}}$ & 164.1125(5) \\ \hline
        \rule{0pt}{3ex}11 & Na &  $\mathrm{3s^1}\,\mathrm{^2S_{1/2}}$  & 162.7(5) \\ \hline
        \rule{0pt}{3ex}12 & Mg &  $\mathrm{3s^2}\,\mathrm{^1S_{0}}$  & 71.2(4) \\ \hline
       \rule{0pt}{3ex}19 & K &$\mathrm{4s^1}\,\mathrm{^2S_{1/2}}$  &289.7(3) \\ \hline
        \rule{0pt}{3ex}20 & Ca & $\mathrm{4s^2}\,\mathrm{^1S_{0}}$  &$160.8(4.0)$\\ \hline
         \rule{0pt}{3ex}24 & Cr & 
         $\mathrm{3d^5}\, \mathrm{^7S_3}$  &$83(12)$\\ \hline
        \rule{0pt}{3ex}37& Rb& $\mathrm{4p^65s^1}\,\mathrm{^2S_{1/2}}$  & 319.8(3) \\ \hline
        \rule{0pt}{3ex}38&Sr & $\mathrm{4p^65s^2}\,\mathrm{^1S_{0}}$  & 197.2(2) \\ \hline
        \rule{0pt}{3ex} 55&Cs &  $\mathrm{4d^{10}5p^66s^1}\,\mathrm{^2S_{1/2}}$ & 400.9(7)\\ \hline
        \rule{0pt}{3ex}70&Yb & $\mathrm{5p^64f^{14}6s^2\,^1S_{0}}$  &139(6) \\ \hline
        \rule{0pt}{3ex}\bf80& \bf Hg& $\mathrm{5d^{10}6s^2}\,\mathbf{^{\!1}S_0} $&\bf 33.91(34)\\ \hline
    \end{tabular}
    \caption{Static scalar dipole polarizabilities, $\alpha_s$, of the elements commonly cooled in combined MOT + ODT experiments. Values (uncertainties in parentheses) are given in atomic units $e^2a_0^2/E_h$, where $e$ is the elementary charge, $E_h$ is the Hartree energy, and $a_0$ is the Bohr radius. Data are taken from the recommended values in Ref.~\cite{Schwerdtfeger2019}.
%    Static scalar dipole polarizability values $\alpha_s$ for the elements most commonly cooled in combined MOT + ODT. The values and their associated uncertainties in the parentheses are expressed in atomic units $a_0^3$, where $a_0$ is Bohr radius.
    %($a.u.=1.6487773\times10^{–41}$~Cm$^2$/V). 
 %   The tabulated values are the recommended values reported in~\cite{Schwerdtfeger2019}.
 }
    \label{tab:polarizabilities}
\end{table}
Table I compares the static scalar dipole polarizability of mercury with those of other commonly trapped atomic species, using the recommended values reported in Ref.~\cite{Schwerdtfeger2019}.
%, emphasizing its exceptionally low value. 

Mercury exhibits low sensitivity to black-body radiation, further enhancing its potential for precision metrology, such as in atomic clocks~\cite{Hachisu2008,Petersen2008,Mcferran2012,Guo2023}. 
Research on mercury spans both the ultracold regime~\cite{Hachisu2008,Liu2013,Lavigne2022} and room temperature vapor studies~\cite{Srivastava2018,Witkowski2019,Linek2022,Gravina2024,Gravina2026}. Interactions between ultracold mercury and other atomic species are also being investigated~\cite{Witkowski2017,Borkowski2017, Bala2026}.
%In addition, studies on interactions between ultracold mercury and other elements are also underway~\cite{Witkowski2017, Bala2026}.

Despite these unique properties, optical dipole trapping of mercury has not yet been demonstrated, primarily because the weak trapping potential requires high optical intensities and efficient loading from a MOT. Realizing an ODT for Hg is therefore considerably more challenging than for most other laser-cooled atomic species.

%Despite these unique properties, optical dipole trapping of mercury has not been demonstrated so far, primarily because the weak trapping potential requires high optical intensities and efficient loading from a MOT. Realizing an ODT for Hg is therefore considerably more demanding than for most other laser-cooled atomic species.
\begin{figure*}
    \centering
    \includegraphics[width=0.8\linewidth]{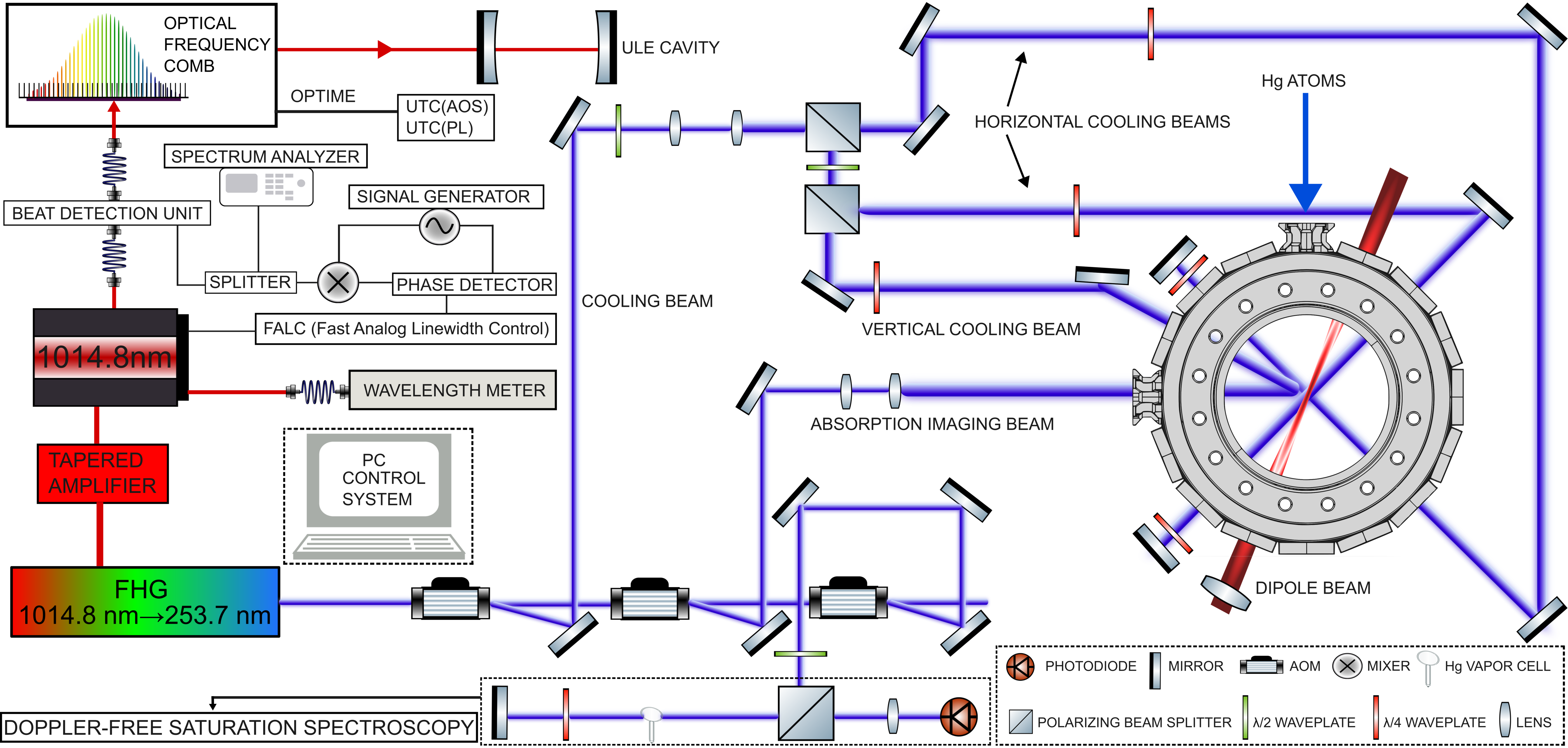}
    \caption{Simplified schematic of the laser system for Hg MOT and ODT. A frequency-quadrupled laser provides up to 100~mW of radiation at 253.7~nm. Three acousto-optic modulators (AOMs) control the laser frequency for MOT cooling, absorption imaging, and Doppler-free saturation spectroscopy. The laser frequency is stabilized to an optical frequency comb referenced to UTC(AOS) and UTC(PL) via the OPTIME stabilized fiber network~\cite{Azoubib03,Jiang15,Morzynski2015,Krehlik15,Sliwczynski13}.}
    %A simplified schematic of the laser setup for the generation of a Hg MOT. A fourth-harmonic generation laser system delivers up to 100 mW of light at 253.7 nm. The output beam from the laser passes through a sequence of AOMs, which shift its frequency as follows: the first AOM red-detunes the beam by 1.4~$\Gamma$ and splits it into three pairs of cooling beams; the second AOM sets the frequency of the imaging beam; and the third AOM adjusts the resonance beam frequency based on a Doppler-free saturation absorption signal.
%A portion of the 1014.8 nm laser light is directed through AOM2 to a transfer cavity. 
%The frequency of the ECDL is stabilized to an optical frequency comb, which is referenced to UTC(AOS) and UTC(PL)~\cite{Azoubib03,Jiang15} via the stabilized fiber-optic link~\cite{Morzynski2015} of the OPTIME network~\cite{Krehlik15,Sliwczynski13}.}
    \label{fig:exp_setup}
\end{figure*}
In this letter, we report the first realization of a single-beam ODT for ultracold mercury atoms. Prior to this work, magnesium~\cite{Riedmann2012} and chromium~\cite{Griesmaier2005}, with a polarizability more than twice that of Hg, represented the least polarizable atomic species successfully confined in an ODT. We demonstrate the transfer of the six most abundant stable Hg isotopes from a MOT into the ODT and characterize the loading dynamics, trap depth, and lifetime.
%We demonstrate the transfer of several Hg isotopes from a MOT into the ODT and characterize the loading dynamics, trap depth, and trap lifetime. 
The demonstrated ability to confine dense samples of ultracold mercury provides a new platform for studies of photoassociation~\cite{Walther2007}, isotope-dependent collision properties, and future searches for new physics beyond the Standard Model~\cite{Borkowski2019}, and constitutes an essential step toward quantum-degenerate mercury gases.

\begin{figure}[h]
    \centering
    \includegraphics[width=0.8\linewidth]{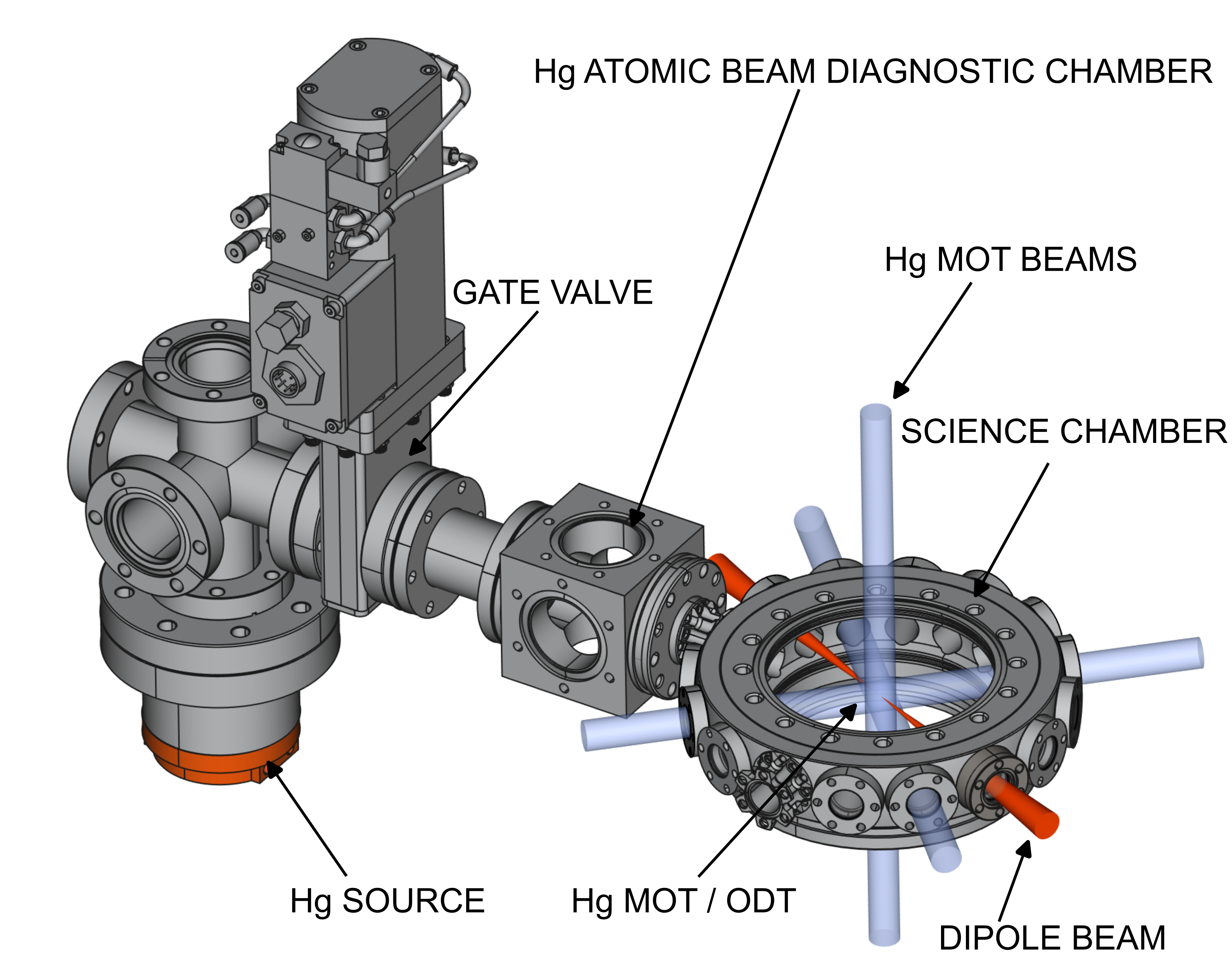}
    \caption{Schematic of the vacuum system used for the preparation of ultracold Hg atoms and their transfer into the ODT. The setup includes a temperature-controlled Hg source, a diagnostic chamber with differential pumping, and the science chamber. The geometry of the MOT cooling beams and the focused ODT beam is also shown.
    %Schematic of the vacuum system used for the preparation of ultracold Hg atoms and their transfer into the optical dipole trap. The temperature-controlled Hg source is connected to the main science chamber through a diagnostic chamber incorporating a differential pumping stage, which maintains a pressure difference of approximately three orders of magnitude while enabling characterization of the atomic beam. The figure also shows the geometry of the MOT cooling beams and the focused dipole beam forming the ODT.
    }
    \label{fig:vac}
\end{figure}

\textit{Experimental system}---The experimental setup has been described in detail in Ref.~\cite{Witkowski2017}; here we summarize only the components relevant to the implementation of the optical dipole trap. Ultracold mercury atoms are first prepared in a MOT operating on the $^1\mathrm{S}_0-{}^3\mathrm{P}_1$ transition at 253.7~nm. A schematic of the experimental setup is shown in Fig.~\ref{fig:exp_setup}, while the vacuum system is presented in Fig.~\ref{fig:vac}. The cooling beam is generated by a frequency-quadrupled laser system referenced to an optical frequency comb~\cite{Witkowski2019}. Long-term stability is ensured via a reference to the frequencies of UTC(AOS) and UTC(PL)~\cite{Azoubib03,Jiang15} via the stabilized fibre optic link~\cite{Morzynski2015} of the OPTIME network~\cite{Krehlik15,Sliwczynski13}.

Three acousto-optic modulators (AOMs) control the laser frequency and distribute the optical power between the MOT, absorption imaging, and saturated-absorption spectroscopy. The MOT consists of three retro-reflected beam pairs with a $1/e^2$ diameter of 10~mm and a magnetic-field gradient of 15~G/cm generated by water-cooled anti-Helmholtz coils. Under typical operating conditions, the science chamber pressure is $5\times10^{-10}$~mbar.

\begin{figure}[h]
    \centering
    \includegraphics[width=1\linewidth]{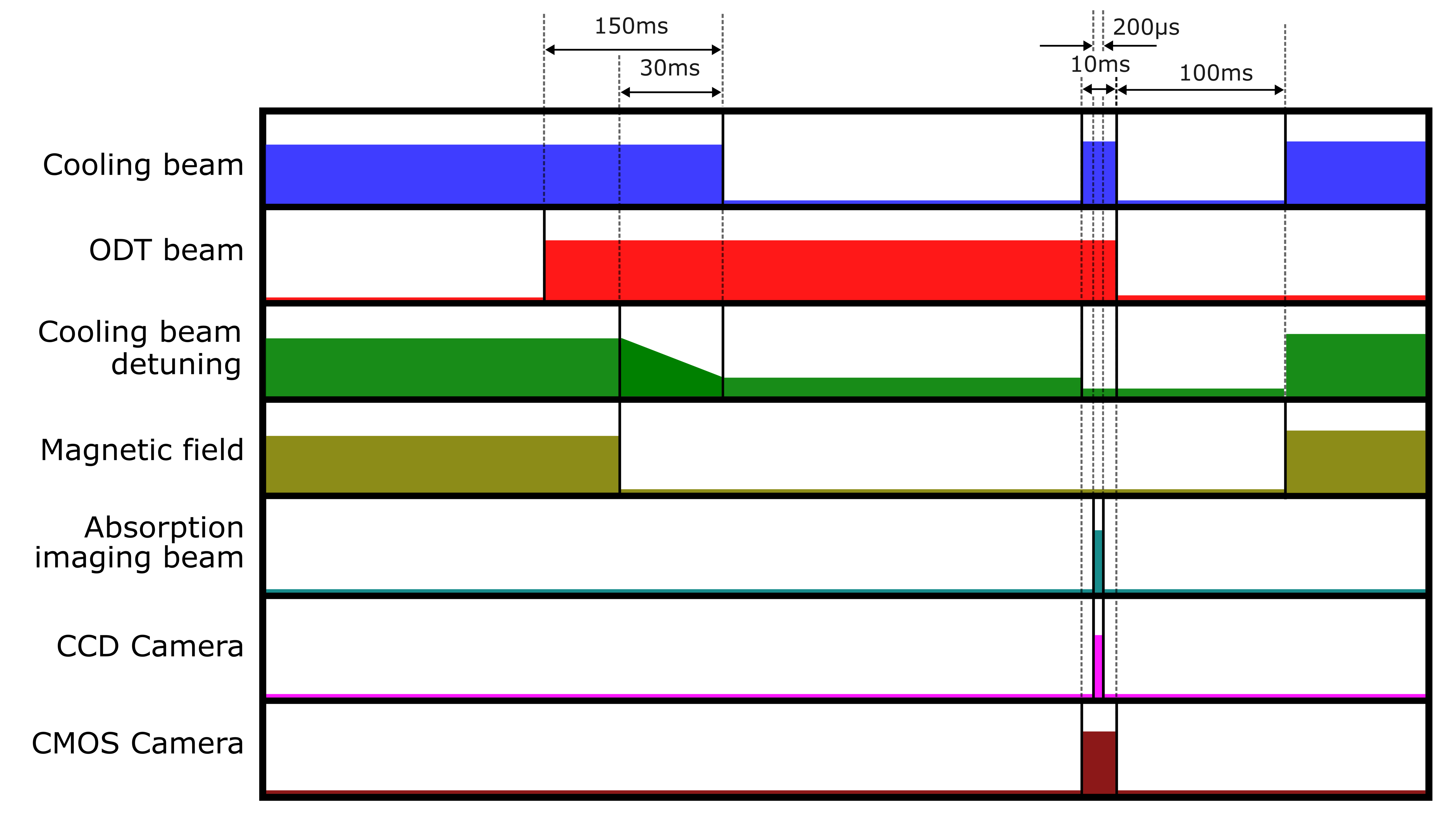}
    \caption{Simplified timing sequence of the experiment. Following MOT loading, the atoms are transferred to a 30~ms optical molasses before being held in the ODT and detected by absorption imaging. For parametric excitation measurements, the ODT intensity is modulated during the holding stage.
    %Simplified timing diagram of the experimental sequence. After the MOT loading stage, the atoms are transferred to an optical molasses lasting 30~ms, during which the cooling laser detuning is linearly ramped from -15~$\Gamma$ to -1.4~$\Gamma$. The detuning is then further reduced to resonance for absorption imaging. For the parametric heating measurements, the intensity of the ODT beam is modulated during the holding stage.
    }
    \label{fig:Time diagram}
\end{figure}

\begin{figure*}
    \centering
    \includegraphics[width=1\textwidth]{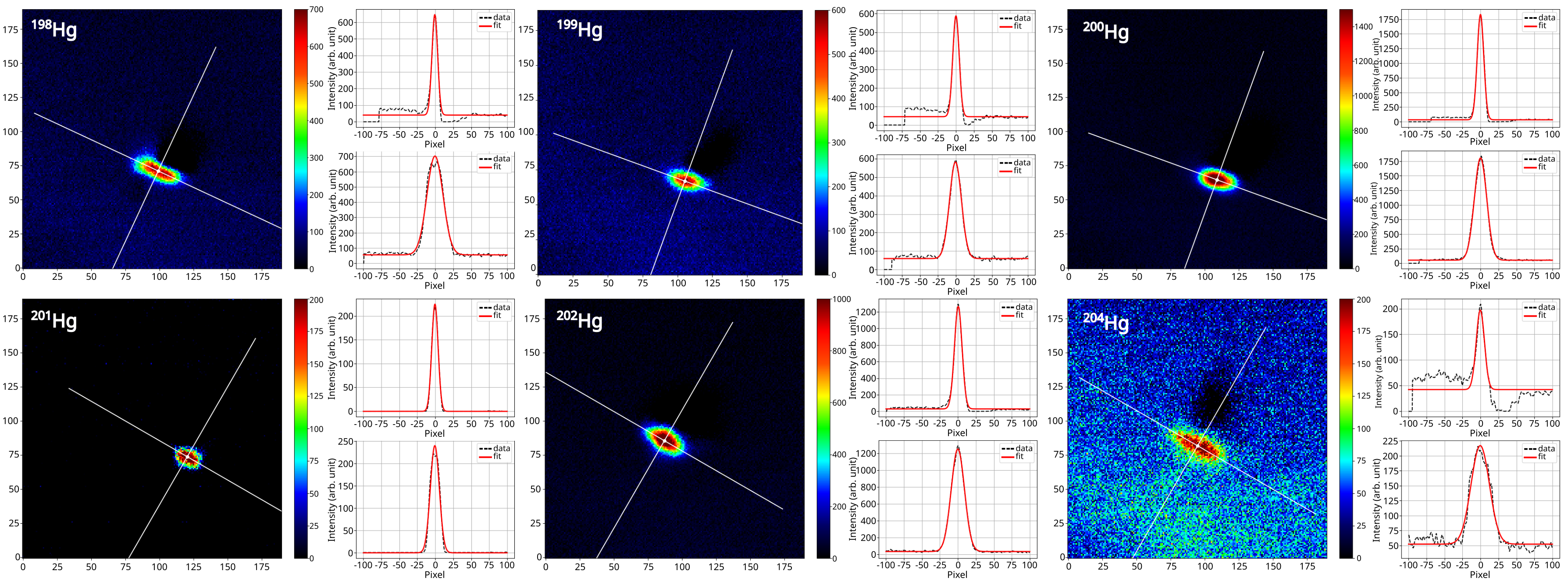}
    \caption{CCD images of the six most abundant stable Hg isotopes after 2~ms of time-of-flight expansion, together with the axial and radial density profiles.}
    \label{fig:wide}
\end{figure*}

The experimental sequence is shown in Fig.~\ref{fig:Time diagram}. Mercury atoms are first loaded into the MOT for approximately 4~s. The ODT is switched on 120~ms before the magnetic field is turned off, followed by a 30~ms optical molasses stage during which the cooling-beam detuning is linearly ramped from -15~$\Gamma$ to -1.4~$\Gamma$, where $\Gamma=1.27$~MHz is the natural linewidth of the cooling transition. At the end of the optical molasses phase, the cooling beams are switched off and their frequency is tuned to resonance. The atoms confined in the ODT are detected by absorption or fluorescence imaging following a variable holding time. Depending on the measurement, the sequence can be modified by releasing the atoms for time-of-flight imaging (examples are shown in Fig.~\ref{fig:wide}) or by applying a modulation to the ODT intensity for parametric excitation measurements.

The optical dipole trap is formed by a single tightly focused laser beam generated by a 300~W continuous-wave ytterbium fiber laser operating at a wavelength of 1070~nm. The laser beam, with an initial diameter of 4.7~mm and a beam quality factor $M^2=1.03$, is focused by a 150-mm-focal-length lens to a waist of approximately $20~\mu$m, corresponding to a Rayleigh length of about 1.2~mm. Although the laser is capable of delivering up to 300~W of optical power, the trap typically operates at powers not exceeding 240~W.
%\section{Optical Dipole Potential}

\textit{Optical Dipole Potential}---The optical dipole potential experienced by an atom in the electronic ground state  
\begin{equation}
     U_{\rm{dip}}(r,z)=-\frac{1}{2\epsilon_0 c}\text{Re}\left(\alpha\left(\omega_{\rm{dip}}\right)\right) I(r,z),
     \label{eq:udip}
 \end{equation}
is proportional to dynamic polarizability of the atom in its ground state $\alpha(\omega_{\rm{dip}})$ and the local laser intensity of the Gassian beam
\begin{equation}
    I(r,z)=\frac{2P}{\pi w^2(z)}\exp\left(-\frac{2r^2}{w^2(z)}\right),
\end{equation}
where $r$ and $z$ denote the radial and axial coordinates, respectively, and $P$ is the total optical power of the trapping beam. 
The beam radius $w(z)$ varies along the propagation direction according to $w(z)=w_0\sqrt{1+\left({z}/{z_R}\right)^2}$
%\begin{equation}
 %   w(z)=w_0\sqrt{1+\left(\frac{z}{z_R}\right)^2},
%\end{equation}
where $w_0$ is the beam waist and $z_R$ is the corresponding Rayleigh length, given by
%\begin{equation}
    %z_R=\frac{\pi w_0^2}{\lambda}=\frac{\pi w_0^2 \nu}{c}.
 $   z_R=\pi w_0^2/\lambda_{\rm{dip}}$.
    %z_R=\frac{\pi w_0^2 \omega}{2\pi c}.
%\end{equation}

Expressing $\alpha(\omega_{\rm{dip}})$ in terms of the oscillator strengths $f_j$ of the electric-dipole transitions to all excited states $j$ yields~\cite{Grimm2000}
 \begin{equation}
     U_{\rm{dip}}(r,z)=-\frac{e^2}{2\epsilon_0 m_e c}I(r,z)\sum_j \frac{f_j}{\omega_j^2-\omega_{\rm{dip}}^2},
     \label{eq:sum}
 \end{equation}
\noindent where $m_e$ denotes the electron mass, $\omega_j$ is the angular frequency of the transition from the ground state to the excited state $j$, and the summation extends over all excited states.
%A complete evaluation of the summation in Eq.~(\ref{eq:sum}) would require oscillator strengths for all excited states, including higher-lying states, for which experimental and theoretical data are incomplete. Therefore, 
We approximate the ground-state polarizability by including only the two strongest contributions from the lowest-lying allowed transitions to the $^1\mathrm{P}_1$ and $^3\mathrm{P}_1$
states. The oscillator strengths of these two transitions account for approximately 60\% of the total static ground-state polarizability~\cite{Romalis1999}. Although this approximation neglects contributions from other excited states, including the $^3\mathrm{P}_0$ and $^3\mathrm{P}_2$ states and higher-lying levels, the dominant $^1\mathrm{P}_1$ and $^3\mathrm{P}_1$ contributions provide a reasonable estimate of the ODT depth.
%Although this approximation neglects the contribution of higher-lying excited states, it provides a reasonable estimate of the ODT depth. 
Using Eq.~(\ref{eq:sum}) and a trapping laser power of 210~W, we obtain a trap depth of approximately 1~mK, which is about 32 times larger than the Doppler temperature of mercury.

\begin{figure}[h!]
    \centering
    \includegraphics[width=0.8\linewidth]{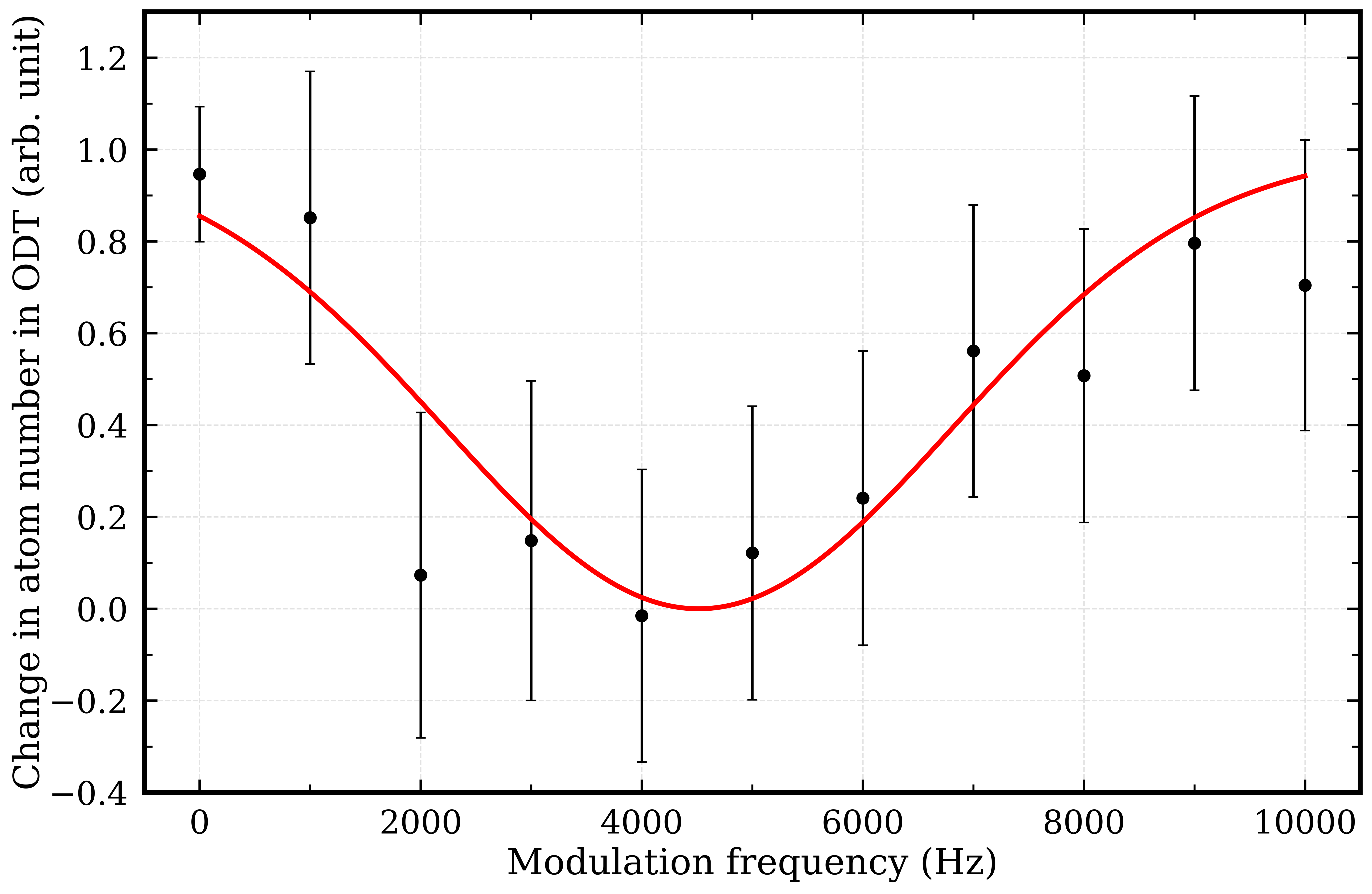}
    \caption{Change in the number of $^{202}$Hg atoms remained in the ODT as a function of the dipole-beam intensity modulation. The red solid line represents a Gaussian fit to the experimental data, yielding a parametric resonance frequency of $\Omega_{\mathrm{pr}}=2\pi\times4.5(5)\,\mathrm{kHz}$. }
   \label{Fig: Freq}
\end{figure}

\textit{Experimental results}---The ODT depth calculated using Eq.~(\ref{eq:sum}) was compared with an experimentally determined value. For this purpose, the radial parametric resonance frequency $\Omega_{\mathrm{pr}}$ was measured by applying a sinusoidal modulation of the ODT beam intensity with a modulation depth of 25\% and monitoring the number of atoms remaining in the trap as a function of the modulation frequency~\cite{Friebel1998}. Fig.~\ref{Fig: Freq} shows the change in the number of atoms remained in the ODT as a function of the dipole-beam intensity modulation frequency. The red solid line represents a Gaussian fit to the experimental data, yielding a parametric resonance frequency of $\Omega_{\mathrm{pr}}=2\pi\times4.5(5)$~kHz, at which the maximum atom loss is observed.
 Due to the low signal level, averaging over a large number of measurements was required to obtain a sufficient signal-to-noise ratio. Each data point represents the average of 2000 independent measurements, and the corresponding uncertainties were determined as the standard error of the mean.

 The radial trap oscillation frequency $\Omega_{0}=\Omega_{{\rm{pr}}}/2$ allows determination of the ODT depth using the relation~\cite{Grimm2000}
 
 %The trap oscillation frequency $\Omega_0$ is determined as one-half of $\Omega_{\mathrm{pr}}$ -- the modulation frequency at which the maximum atom loss is observed.
 %Once the $\Omega_0$ has been determined, the ODT depth can be calculated using the relation
\begin{equation}
    U_{\mathrm{dip}}=-\frac{1}{4}m\Omega_0^2 w_0^2,
    \label{eq:u_dip_resonance}
\end{equation}
where $m$ is the Hg mass.
The ODT depth determined from the measured value of $\Omega_{\mathrm{pr}}$ using Eq.~(\ref{eq:u_dip_resonance}) is $\left|U_{\rm{dip}}\right|/k_{B}=0.61(14)$~mK, where $k_{B}$ is Boltzmann's constant. 
The value obtained in this way differs from the value 1~mK estimated based on Eq.~(\ref{eq:sum}). This discrepancy may result from the oversimplified model of the dipole beam intensity distribution and from the harmonic approximation of the trapping potential, which neglects anharmonic effects. Such effects can become significant for shallow optical traps.

\begin{figure}
    \centering
    \includegraphics[width=0.47\linewidth]{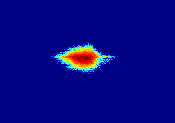}
    \includegraphics[width=0.51\linewidth]{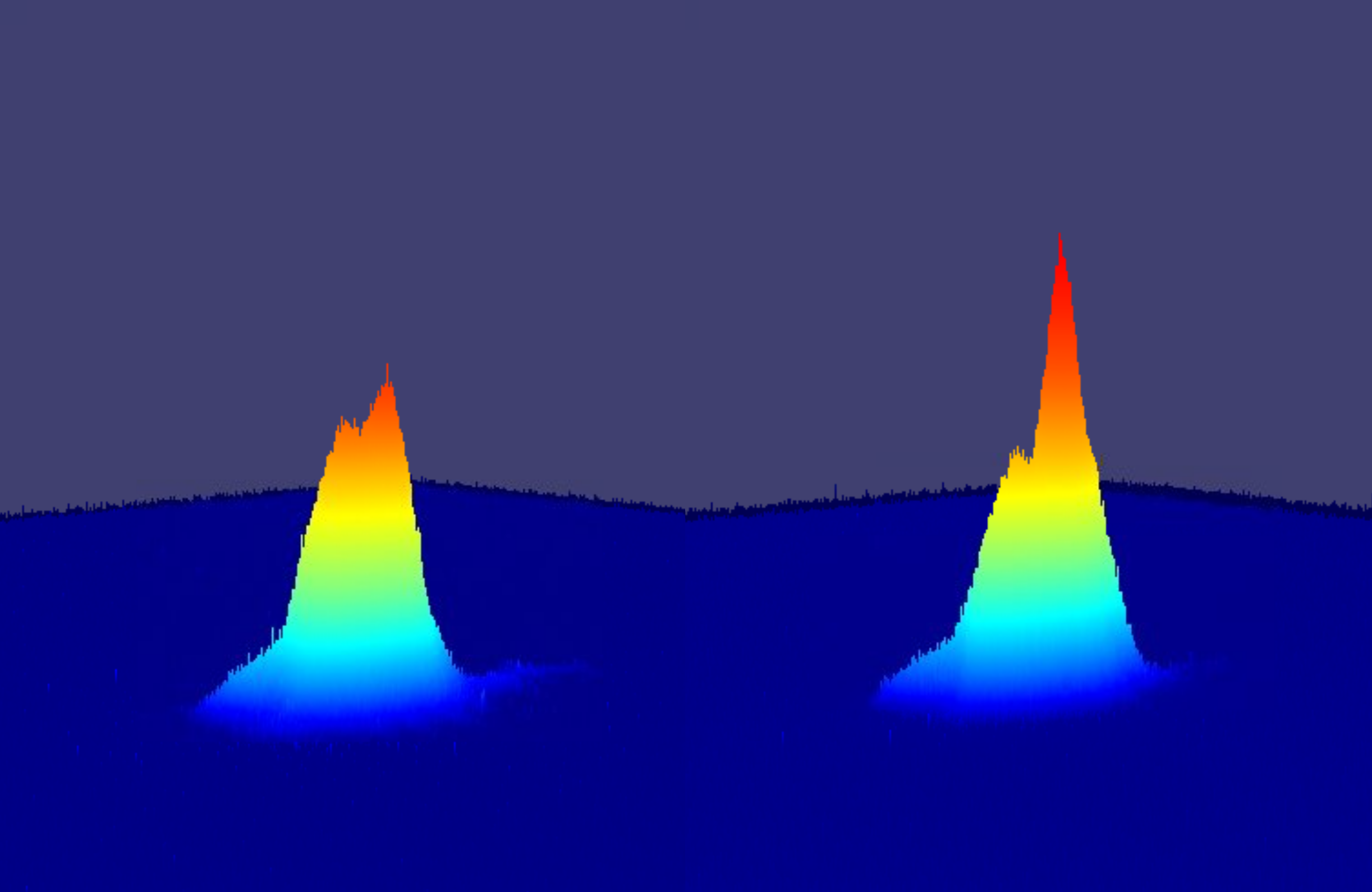}
    \caption{Left: Image of the $^{202}$Hg MOT with the ODT overlapped, recorded using a CMOS camera.  Right: Three-dimensional intensity profiles of the $^{202}$Hg MOT with and without the ODT. The centers of the MOT and ODT are slightly displaced with respect to each other; this configuration corresponds to the optimal transfer of atoms into the ODT.}
    \label{fig:3d}
\end{figure}
\begin{figure}
    \centering
    \includegraphics[width=0.8\linewidth]{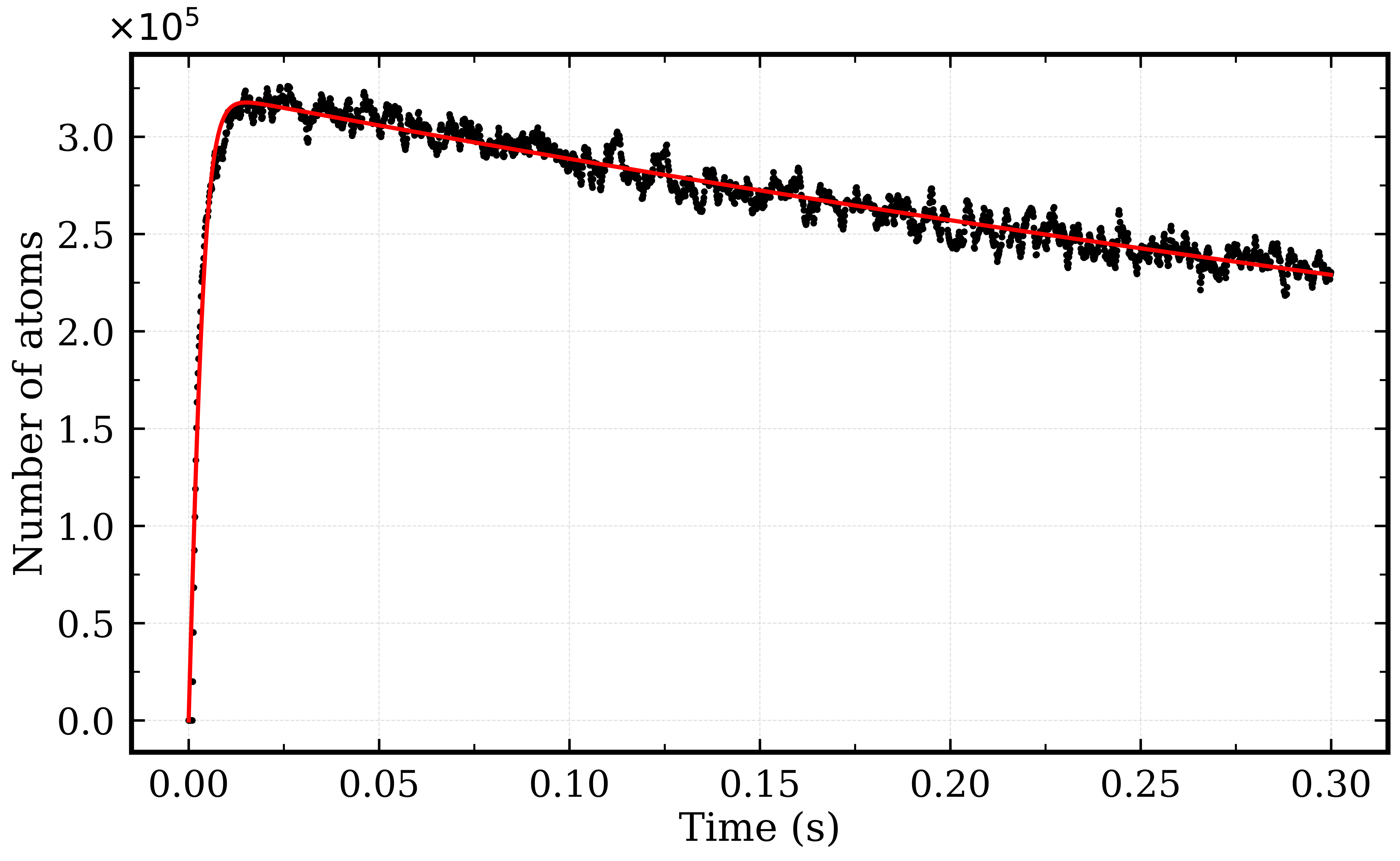}
    \caption{Loading curve of $^{202}$Hg atoms into the ODT in the presence of the MOT cooling beams. The red solid line represents a fit using Eq.~(\ref{eq:mot_loading}).
    %, yielding the parameters $\gamma=2.312(7)\,\mathrm{s^{-1}}$, $\beta=68.29(57)\,\mathrm{s^{-1}}$, $\Gamma=1.53(21)\,\mathrm{s^{-1}}$, and $L_0=718(6)\,\mathrm{s^{-1}}$. 
    The maximum atom number of $3.176(2)\times10^5$ is reached after approximately~15 ms of loading.
    }
    %Number of atoms loaded in the FORT vs FORT loading stage duration%\ %$\Gamma= 7.186622e-9 \pm 0.1327$ 
    %\ $\beta= 66.3(46)$
 %\ $\gamma= 2.36(13)$, Trap parameters: $\omega_0=20\mu m$ }
    \label{fig:loading}
\end{figure}
To further characterize the transfer process and identify the factors limiting the number of atoms transferred to the ODT, we analyze the loading dynamics from the MOT into the ODT.
%To further characterize the loading process and identify the factors limiting the number of atoms transferred into the ODT, we analyzed the loading dynamics from the MOT into the ODT.
The ODT loading dynamics are determined by the interplay between the loading rate and atom loss processes.
%The transfer of atoms from a MOT to an ODT is governed by a competition between the loading rate and atom loss mechanisms.
Efficient loading requires good phase-space matching between the atomic cloud and the trapping potential. We observed that the maximum number of atoms transferred into the ODT is achieved when the trap is slightly offset from the center of the MOT rather than being perfectly overlapped with it, as shown in Fig.~\ref{fig:3d}. This observation is consistent with the theoretical predictions presented in Ref.~\cite{Szczepkowicz2009}. %\textcolor{blue}{Early experiments achieved the transfer of approximately 1300 optically molasses-cooled atoms into an ODT~\cite{Miller1993}, while later studies demonstrated that continuous overlap of the ODT with the MOT can increase the number of trapped atoms to as many as~\cite{Adams1995,Boiron1996}.\bf I am not sure if this historical background should stay here.} 
Additional images showing the spatial distributions of atoms released from the ODT after 2~ms of ballistic expansion for the investigated Hg isotopes are presented in Fig.~\ref{fig:wide}. The temperature of the $^{202}$Hg atomic cloud after release from the ODT is estimated from time-of-flight expansion to be below 0.1~mK. 

\begin{figure}[h]
    \centering
    \includegraphics[width=0.8\linewidth]{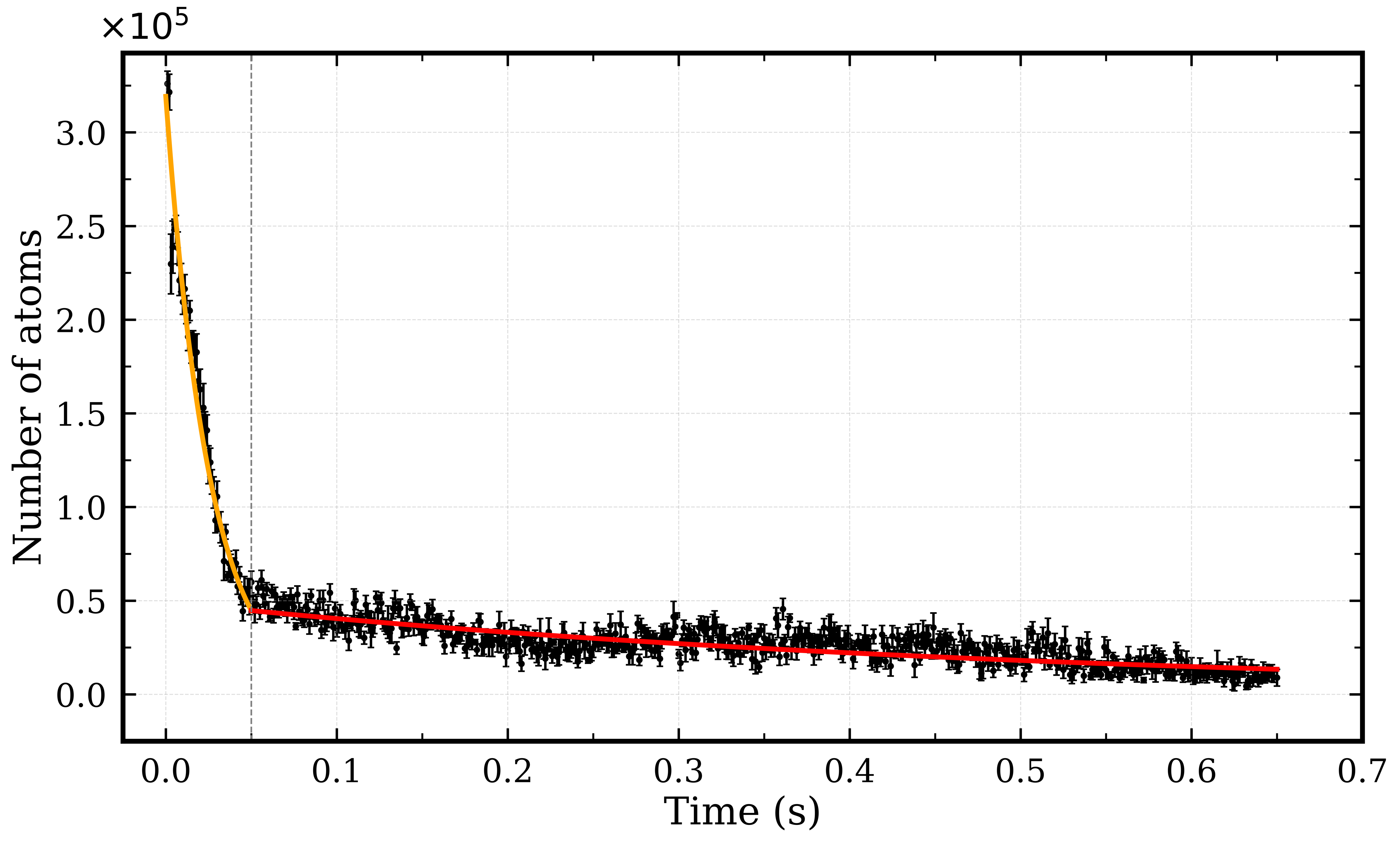} 
    \caption{Decay of the number of $^{202}$Hg atoms remaining in the ODT after the magnetic field is switched off. The decay is well described by Eq.~(\ref{eq:two_component_decay}) and consists of distinct fast and slow components. The two regimes are separated by the vertical gray dashed line. The corresponding fits to the fast and slow decay components are shown by the orange and red solid lines, respectively.}

    \label{Fig:lifetime}
\end{figure}
In our experiment, the loading dynamics were investigated by measuring the number of atoms trapped in the ODT as a function of the loading time. 
A representative ODT loading curve for $^{202}$Hg in the presence of the cooling beams is shown in Fig.~\ref{fig:loading}.  The atom number initially increases rapidly, reaching a maximum after approximately 15~ms, corresponding to 3.176(2)$\times$10$^{5}$ trapped atoms and a peak density of 0.50(7)$\times$10$^{12}$~cm$^{-3}$. At longer loading times, the atom number gradually decreases. The loading dynamics are described by~\cite{Kuppens2000}
\begin{equation}
\frac{dN}{dt}=L_{0}e^{-\gamma t}
-
\kappa N
-
\beta N^{2},
\label{eq:mot_loading}
\end{equation}
where $N$ is the number of atoms trapped in the ODT, $L_0$ is the initial loading rate, $\gamma$ characterizes the gradual reduction of the loading rate resulting from the decreasing efficiency of magneto-optical cooling within the ODT region, $\kappa$ is the one-body loss coefficient associated with collisions with the background gas, and $\beta$ is the two-body loss coefficient arising from cold atom collisions. 
The reduction of the cooling efficiency may originate from the light shift induced by the ODT. Its magnitude is difficult to determine precisely because the polarizability of Hg in the excited state is not known with sufficient accuracy. If we assume that the optical potential depth for Hg in the $^3\rm{P}_1$ state is twice as large as that in the $^1\rm{S}_0$ state, the estimated light shift does not exceed 20~MHz.
%If we would assume that the optical potential depth for Hg in the $^3P_1$ state is two times larger than that in the $^1S_0$ state, estimated light shift does not exceed 20~MHz. 
This estimate is based on the experimentally determined maximum detuning at which atoms can be trapped in the MOT. 
%Since the fraction of atoms in the excited state in our MOT does not exceed 20\% under our experimental conditions, the dominant contribution is expected to arise from the ground-state polarizability, resulting an estimated light shift of no more than 10 MHz. 
Such a shift would increase the effective detuning of the MOT cooling light within the ODT volume, thereby reducing the cooling efficiency and, consequently, the loading rate into the ODT. Similar behavior has been reported in previous experiments~\cite{Kuppens2000,Sofikitis2011}.
The loading dynamics are qualitatively similar for all investigated isotopes.  The loading curve shown in Fig.~\ref{fig:loading} corresponds to the $^{202}$Hg isotope and is fitted with $\gamma=2.312(7)$~s$^{-1}$, $\beta=6.829(8)\times10^{-4}$~s$^{-1}$, $\kappa=1.5(2)$~s$^{-1}$, $L_0=7.18(2)\times10^7$~s$^{-1}$.

To determine the ODT lifetime, the number of atoms remaining in the trap was measured as a function of the time after the magnetic field was switched off. The decay curve obtained for $^{202}$Hg is shown in Fig.~\ref{Fig:lifetime}. The observed decay deviates from a single-exponential form and is instead described by two distinct decay components
\begin{equation}
N(t)=A_{1}e^{-\Gamma_{\mathrm{fast}}t}
+A_{2}e^{-\Gamma_{\mathrm{slow}}t}.
\label{eq:two_component_decay}
\end{equation}
An initial fast decay, lasting approximately the first 50~ms, is characterized by the decay rate $\Gamma_{\mathrm{fast}}$ = 39.3(4)~s$^{-1}$. At longer times, the atom number decays more slowly, with a decay rate of $\Gamma_{\mathrm{slow}}=2.00(3)$~s$^{-1}$. The ODT lifetime is determined from the slower decay component and is given by $1/\Gamma_{\mathrm{slow}}=498.2(8)$~ms. 

The rapid initial decay is most likely associated with density-dependent collisional loss processes, such as three-body recombination or photoassociation-induced losses, since the MOT beams remain on for the first 
30~ms after the magnetic field is switched off. These processes are expected to dominate at high atomic densities. 
%The rapid initial decay is most likely associated with density-dependent collisional loss processes, such as three-body recombination or photoassociative losses (MOT beam are present in first 30~ms after switching off the magnetic field), which are expected to dominate at the high atomic densities. 
Since these processes depend on the scattering length, their magnitude is expected to vary for different isotope pairs. This dependence opens a promising route for future measurements of isotope-dependent scattering properties using ODT lifetime measurements. However, such measurements are out of the scope of this letter.

%\section{Conclusion}
\textit{Conclusion}---We have demonstrated the first realization of an ODT for ultracold mercury atoms and characterized the transfer of atoms from a MOT into the dipole potential. Despite the low polarizability of Hg, which makes optical confinement challenging, we achieved stable trapping of dense atomic samples and investigated the loading dynamics using a rate-equation model. The ODT depth was independently determined through measurements of the radial parametric oscillation frequency, providing an experimental characterization of the trapping potential. The decay dynamics revealed two distinct loss components, with the long-time decay determining the ODT lifetime.

The demonstrated ability to prepare dense ultracold mercury samples in an ODT provides a new platform for experiments where interatomic interactions play a key role. In particular, the presented system enables future studies of isotope-dependent collision properties, photoassociation processes, and other density-dependent phenomena in ultracold mercury. This work establishes the experimental foundation for further investigations of Hg-based quantum systems and precision measurements.

Finally, our work opens the possibility of optical dipole trapping of other atomic systems with polarizabilities comparable to that of Hg. These include species already laser cooled and trapped in a MOT, such as Ag~\cite{Uhlenberg2000}, Cd~\cite{Brickman2007}, Zn~\cite{Moller2025}, as well as systems that remain to be laser cooled, such as Al, Si, Cu, Ni, Xe. This is also a step towards optical dipole trapping of H, H$_2$ and He.

\textit{Note added}---During the preparation of this manuscript, we became aware of independent efforts toward the realization of an optical dipole trap for mercury atoms by the group of Prof. Simon Stellmer at the University of Bonn~\cite{Groh2025}.

\textit{Data Availability}---The datasets generated and analyzed during the current study are available in the open repository~\cite{repod}.

%\section{Acknowledgements}
\textit{Acknowledgements}---I. Nandi and R. Ciury\l{}o acknowledge Polish National Science Centre Project No. 2021/41/B/ST2/00681 support. A. Sahu, M. Veis and M. Witkowski acknowledge Polish National Science Centre Project No. 2021/42/E/ST2/00046 support. 

The research was performed at the National Laboratory FAMO (KL FAMO) in Toru\'n, Poland, and was supported by a subsidy from the Polish Ministry of Science and Higher Education.
%\section{Funding}

%\textit{Funding}---I. Nandi and R. Ciury\l{}o acknowledge Polish National Science Centre Project No. 2021/41/B/ST2/00681 support. A. Sahu, M. Veis and M. Witkowski acknowledge Polish National Science Centre Project No. 2021/42/E/ST2/00046 support. 

%\bibliographystyle{unsrt}
%\bibliography{sample}
\bibliographystyle{apsrev4-2}
\bibliography{sample}

\end{document}